\documentclass[%
superscriptaddress,
nofootinbib,
amsmath,
amssymb,
prb,
twocolumn
]{revtex4-2}

\usepackage{dsfont}
\usepackage{bbm}
\usepackage{graphicx}
\usepackage{dcolumn}
\usepackage{bm}
\usepackage{xcolor} 
{\color{red}} 
{} 

\begin{document}

\preprint{APS/123-QED}

\title{\textbf{Finite-Size Effects on Metallization vs. Chiral Majorana Fermions} 
}%

\author{Xin Yue}
\affiliation{Beijing Computational Science Research Center, Beijing 100193, China} 
\author{Guo-Jian Qiao}%
\email{qiaogj0211@csrc.ac.cn}
\affiliation{Beijing Computational Science Research Center, Beijing 100193, China}
\affiliation{Graduate School of China Academy of Engineering Physics, Beijing 100193, China}
\author{C. P. Sun}
\email{suncp@gscaep.ac.cn}
\affiliation{Graduate School of China Academy of Engineering Physics, Beijing 100193, China}

\begin{abstract}
The search for chiral Majorana fermions in quantum anomalous Hall insulator/\textit{s}-wave superconductor heterostructures has attracted intense interest, yet remains controversial due to the lack of conclusive evidence. A key issue is that the heterostructure's metallization can produce half-integer conductance signatures resembling those of chiral Majorana fermions, thereby complicating their identification. In this Letter, we investigate how the competition between metallization and chiral Majorana fermions depends on superconductor thickness, revealing its critical role through three distinct regimes: (i) For thin superconductors ($\sim$10 nm), metallization shows periodic oscillations with thickness, matching the Fermi wavelength. (ii) Intermediate thicknesses ($\sim$100 nm) exhibit periodic windows for observing chiral Majorana fermions. (iii) Thick superconductors ($\sim$1000 nm) sustain stable chiral Majorana fermions that are insensitive to thickness variations. These results suggest that superconductor thickness is a key control parameter for advancing efforts to conclusively identify chiral Majorana fermions.\end{abstract}

\maketitle 
 

\emph{Introduction}--The pursuit of Majorana fermions in condensed matter systems over the past two decades has been driven by their potential applications in topological quantum computing \citep{Kitaev_2003, Alicea_2012, beenakker2013, flensberg2021,Li2014,PNAS}. Although theoretical proposals for chiral Majorana fermions in heterostructures composed of quantum anomalous Hall (QAH) insulators proximity-coupled \textit{s}-wave superconductors (SCs) have long been proposed \citep{Read&Green2000,Fu_liang_2008,Qi_2010,Chung_2011, Wang_2015}, experiments about chiral Majorana fermions remains contentious \citep{He2017chiral, DavidAbergelComment2020, KTlaw2017, Huang2018, Wen2018, Huang_2024,Uday2024, 2025PRB}. It was found that metallization effects (SC layer makes the QAH insulator metallic) can produce spurious signatures, which further hinder the identification of chiral Majorana fermions \citep{2020Absence, KTlaw2017, Wen2018, Huang_2024}.

Despite extensive studies on proximity effects---either from the low-energy effective perspective \cite{2010Proximity,Potter&Lee2011,2017lowEnergy_renormalization} or perturbative approaches \citep{Qiao_2022,Yue2023}---none have incorporated metallization into the theoretical framework. Besides, studies of metallization remain phenomenological discussion \citep{KTlaw2017,Huang2018,Wen2018}, lacking microscopic Hamiltonian-level derivations. This leaves a critical gap in unifying metallization and Majorana physics within a single theoretical framework, hindering the ability to provide mechanistic explanations or predictive guidance for avoiding metallization in observing chiral Majorana fermions.

In addition, the thickness of the SC layers used in experiments ranges from 50\,nm to 350\,nm \citep{He2017chiral, Huang_2024, 2025PRB}, which falls within the mesoscopic scale—an intermediate regime between two-dimensional (2D) and three-dimensional (3D) systems. To our best knowledge, this finite-thickness effects of SC \citep{2017_Finite} on the experimental observation of chiral Majorana fermions has not yet been considered. Thus, it is essential to account for this effect and offer precise guidance on choosing the superconducting layer thickness for the observation of chiral Majorana fermions.

\begin{figure}[t]
    \centering
\includegraphics[width=0.7\linewidth]{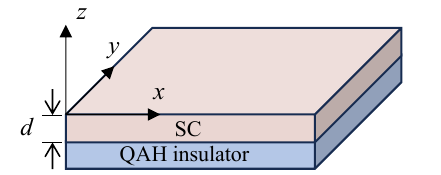}
\maketitle 
    \caption{Heterostructure composed of a superconductor (SC) layer with thickness $d$ layered atop a Quantum Anomalous Hall (QAH) insulator layer.}
    \label{fig:enter-label}
\end{figure}

In this Letter, we utilize a minimal microscopic model of a QAH insulator coupled to a 2D superconductor to explore the effects of proximity coupling strength on chiral Majorana fermions. We introduce the concept of dressed chiral Majorana fermions by extending the theoretical framework of dressed Majorana fermions from 1D hybrid systems~\cite{Qiao2024,zhang2025poor} to our 2D case. The dressed chiral Majorana fermions incorporate excitations of both the QAH insulator and the SC. We demonstrate that metallization arises from a characteristic chemical potential shift combined with a nearly vanishing induced gap. Crucially, for realistic material parameters, the metallization region overlaps with the domain hosting $\mathcal{N}=1$ dressed chiral Majorana fermions, posing a significant challenge for experimental observation.

Incorporating the finite-thickness effect of SC into our analysis, we reveal three main findings: (i) For thin SCs ($\sim$10 nm), there is a periodic structure of metallization with a period corresponding to the SC's Fermi wavelength, and metallization typically occurs except near resonance thicknesses. The periodicity originates from the fact that band shifts and proximity-induced gaps oscillate with the SC's thickness. This oscillatory behavior was also reported in semiconductor nanowire/SC systems~\cite{2017_Finite, Metallization2018}, highlighting the importance of precise thickness control. (ii) For intermediate thicknesses ($\sim$100 nm), the window width for observing chiral Majorana fermion signals exhibits periodic behavior, oscillating at the Fermi wavelength. The observation window is significantly widened when the thickness reaches resonance points. (iii) For thick SCs ($\sim$1000 nm), the behavior of chiral Majorana fermions becomes uniform as the thickness varies, with the proximity-induced gap remaining stable. While thicker SCs generally enhance the proximity effect, precise thickness control for intermediate thicknesses near resonance points can dramatically increase the induced gap, effectively suppress metallization, and significantly widen the observation window for chiral Majorana fermions.

\vspace{0.5em}
\emph{Majorana in microscopic model}--
The heterostructure consisting of a QAH insulator (e.g., Cr- or Fe-doped $\mathrm{Bi}_2\mathrm{Se}_3$) proximitized by a superconductor is commonly described by the phenomenological Bogoliubov-de Gennes Hamiltonian \cite{Qi_2010, Chung_2011}:
\begin{equation}
\mathcal{H}_{\text{P}}(\mathbf{k}) = \begin{pmatrix}
h_{\mathrm{Q}}(\mathbf{k}) -\mu & i \Delta \sigma_y \\
-i \Delta \sigma_y & -h_{\mathrm{Q}}^*(-\mathbf{k})+\mu
\end{pmatrix},
\end{equation}
where the QAH insulator is modeled by the two-band effective Hamiltonian:
\begin{equation}
h_{\mathrm{Q}}(\mathbf{k}) = A(k_{x} \sigma_x + k_{y} \sigma_y)+(m + B k^2) \sigma_z, \label{eq:h_QAH}
\end{equation}
with $A$ denoting the Fermi velocity, $m$ the mass gap parameter, $B$ the band curvature coefficient, and $\sigma_{x,y,z}$ the Pauli matrices. Here $\mu$ represents the chemical potential and $\Delta$ the superconducting pairing amplitude. Chiral Majorana edge modes were predicted based on this phenomenological approach  \cite{Qi_2010}.

Here, we directly solve the holistical microscopic Hamiltonian of the QAH-SC coupled system, $H = H_{\mathrm{Q}} + H_{\mathrm{SC}} + H_{\mathrm{T}}$, to obtain the holistical edge state for chiral Majorana fermions. The QAH insulator component is described as:

\begin{equation}
H_{\mathrm{Q}} = \int \frac{d^2k}{(2\pi)^2} \: \boldsymbol{\varphi}_{\mathbf{k}}^\dagger 
[h_{\mathrm{Q}}(\mathbf{k})-\mu_m]
\boldsymbol{\varphi}_{\mathbf{k}}.
\label{eq:H_QAH}
\end{equation}
Here, $\boldsymbol{\varphi}_{\mathbf{k}} = [\varphi_{\mathbf{k}\uparrow}, \varphi_{\mathbf{k}\downarrow}]^T$, and $\mu_m$ is the chemical potential of QAH insulator. The \textit{s}-wave SC is described by Bardeen-Cooper-Schrieffer (BCS) Hamiltonian:
\begin{equation}
H_{\mathrm{SC}} = \int \frac{d^2k}{(2\pi)^2} \: \mathbf{c}_{\mathbf{k}}^\dagger 
[\epsilon_s \sigma_z + \Delta_s \sigma_x ] 
\mathbf{c}_{\mathbf{k}},
\label{eq:SC}
\end{equation}
where $\mathbf{c}_{\mathbf{k}} = [c_{\mathbf{k}\uparrow}, c_{-\mathbf{k}\downarrow}^\dagger]^T$ represents the Nambu spinor, $\epsilon_s = \hbar^2 \mathbf{k}^{2}/(2 m_s) - \mu_s$ is the kinetic energy above the Fermi level $\mu_s$, $m_s$ is the electron mass and $\Delta_s$ is the superconducting gap. The tunneling coupling between the QAH insulator and SC is:
\begin{equation}
H_{\mathrm{T}} = T \int \frac{d^2k}{(2\pi)^2} \sum_{\sigma=\uparrow,\downarrow} 
\left[ \varphi_{\mathbf{k}\sigma}^\dagger c_{\mathbf{k}\sigma}  + \text{H.c.}\right].
\label{eq:H_T}
\end{equation}
Here, we consider that the proximity coupling strength $T$ is momentum independent and spin conservation is preserved during the tunneling process. Combining Eqs. (\ref{eq:H_QAH}, \ref{eq:SC}, \ref{eq:H_T}), the holistical Hamiltonian in the Nambu basis of $[\boldsymbol{\varphi}_{\mathbf{k}},\boldsymbol{c}_{\mathbf{k}},\boldsymbol{\varphi}_{-\mathbf{k}}^{\dagger},\boldsymbol{c}_{-\mathbf{k}}^{\dagger}]^{T}$, is reexpressed as
\begin{equation}
\mathcal{H}(\mathbf{k}) = 
\begin{pmatrix}
h(\mathbf{k}) & p \\
p^\dagger & -h^*(-\mathbf{k})
\end{pmatrix},
\,
h = 
\begin{pmatrix}
h_{\mathrm{Q}}-\mu_m & T  \\
T  & \epsilon_s 
\end{pmatrix}.
\label{Holistical_BdG}
\end{equation}
Here, $p = \mathrm{diag}(\mathbf{0}, i\Delta_s \sigma_y)$ is a diagonal block matrix containing superconducting pairing.

To obtain the holistical edge state of chiral Majorana fermions in the spatial representation, we now solve the eigen-states of the holistic Hamiltonian with the periodic boundary condition in the $x$ direction and an open boundary condition in the $y$ direction. In this case, $k_y$ is replaced by $-i\partial_y$, and the eigen equation becomes: $
    \mathcal{H}(k_x,-i\partial_y)\mathbf{\Psi}_{k_x}(y)=E(k_x) \mathbf{\Psi}_{k_x}(y)$, where $\mathbf{\Psi}_{k_x}(y)=[\bm{u}_{k_x} (y), \bm{v}_{k_x} (y)]^T $. The corresponding Bogoliubov quasi-excitaion is
\begin{equation}
    \gamma^{\dagger}_{k_x} = \int dy \,[\bm{\psi}_{k_x}^{\dagger}(y) \cdot \bm{u}_{k_x}(y) +  \bm{\psi}_{-k_x}^{T} (y)\cdot \bm{v}_{k_x}(y)],
    \label{quasi}
\end{equation}
where $\bm{o}_{k_x} = [o_{m\uparrow},o_{m\downarrow}, o_{s\uparrow}, o_{s\downarrow}]^T$ with $o(=u,v)$ respectively denotes eigen-wave function of the electron and hole, and $\bm{\psi}_{k_x}(y)$ is the Fourier transformation of $\bm{\psi}_{\mathbf{k}}=[\bm{\varphi}_{\mathbf{k}},\bm{c}_{\mathbf{k}}]^{T}$ in $y$ component. It follows from Eq.~(\ref{quasi}) that the operator that creates a quasiparticle at position $x$ is $\gamma^{\dagger}(x) = \int dk_{x} \,  \gamma_{k_x}^{\dagger}\exp(-i k_x x)$. Following the definition of the Majorana particle (its own antiparticle is itself), the field operator generates the Majorana mode if it satisfies $\gamma(x) = \gamma^{\dagger}(x)$. We name it a dressed chiral Majorana fermion since it includes both electronic excitations from the QAH insulator and quasiexcitations from the SC \cite{Qiao2024}. By the definition of dressed chiral Majorana fermion, we have $\gamma_{k_x}^\dagger = \gamma_{-k_x}$, and further this condition requires that \(\bm{u}_{k_x}(y) = \bm{v}_{-k_x}^*(y)\). 

Assume that the edge state of the dressed chiral Majorana fermion propagates along the $x$-direction, while being localized at the edge in the $y$-direction: $\mathbf{\Psi}_{k_x}(y)=
\exp({-\xi y})[\bm{u}_{k_x},\bm{u}^{*}_{-k_x}]^T$. First, we consider the scenario where \( E = 0 \) and \( k_{x} = 0 \).  In this case, a non-trivial solution exists if $\text{det}[\mathcal{H}(0,i\xi)] = \sum_{n=0}^{16} c_{n} \xi^{n} = 0$. Here, the coefficient \( c_{0} \) is given by
\(
c_{0} = \left[E_s^2(\mu_{m}^{2} - m^{2}) + T^{4} - 2T^{2}\mu_{m}\mu_{s}\right]^2,
\)
where \(E_s^2 = \Delta_{s}^{2} + \mu_{s}^{2}\). The remaining coefficients are not relevant to the primary focus of our analysis. Since \(\xi = 0\) corresponds to an infinite decay length, this condition \( \text{det}[\mathcal{H}(0,0)] = c_0 = 0 \) determines the phase boundary at which the holistic edge mode emerges or disappears. This phase boundary is obtained as:
\begin{equation}
m^{2} = \Delta_{\text{eff}}^{2} + \mu_{\text{eff}}^{2}
\label{eq:Topo_condition}
\end{equation}
with \(\mu_{\text{eff}} = \mu_m + \mathrm{Re}(\chi)\) and \(\Delta_{\text{eff}} = \mathrm{Im}(\chi) \), respectively. Here, \( \chi = -T^2/(\mu_s+i\Delta_s)\) characterizes the dressed effect of the SC on the QAH insulator. 

\vspace{0.5em}
\emph{The number of Majorana edge states in three parameter regions}--
We consider a system of length \( L_y \) in the \( y \)-direction, with the wave function vanishing at the boundaries as required by physical constraints \cite{Shen_PRL_2008}: $ \bm{\Psi}(y=0) = \bm{\Psi}(y=L_y) =0$. For the specific case where $\mu_{\text{eff}}=0$ and under the approximation that $ \mu_s +\hbar^2\xi^2/(2m_s)\approx\mu_s$, the state number of dressed chiral Majorana fermions is determined in three different regions of the parameter space $(m, \Delta_{\text{eff}})$: (i) when $m>\Delta_{\text{eff}}$, no edge state exists; (ii) when \( |m| < \Delta_{\text{eff}} \), there is only one edge state on each edge. These edge states are given by:

\begin{equation}
\begin{aligned}
    \bm{\Psi}_{r,1}(y) &\propto 
    \begin{pmatrix} 
        \bm{u}_{+} \\ 
        \bm{u}_{+} 
    \end{pmatrix}
\left( e^{\xi_{+,1}(y-L_y)} - e^{\xi_{-,1}(y-L_y)} \right)  , \\
    \bm{\Psi}_{l,1}(y) &\propto i 
    \begin{pmatrix}  
        \bm{u}_{-} \\ 
        - \bm{u}_{-} 
    \end{pmatrix}
      (e^{-\xi_{+,1}y} - e^{-\xi_{-,1}y}).
\label{edge_state}
\end{aligned}
\end{equation}
Here, \(\bm{u}_{\pm} = 
\begin{bmatrix}
 \mp E_s^2, &
E_s^2, &
T(\Delta_s \mp \mu_s), &
T( \mu_s \pm\Delta_s  )
\end{bmatrix}^T\) are the coefficient vectors and \(
    \xi_{\pm,1} = \left( A \pm Q \right)/2B\) are the decay factors with \(Q=\sqrt{A^2 - 4B(\Delta_{\text{eff}} - m)}
\); The solutions \(\bm{\Psi}_{r,1}(y)\) and \(\bm{\Psi}_{l,1}(y)\) are invalid when \(m > \Delta_{\text{eff}}\) because, in this case, the decay factors \(\xi_{+,1}\) and \(\xi_{-,1}\) have opposite signs, making it impossible to simultaneously satisfy the boundary conditions at both edges. (iii) when \( m < -\Delta_{\text{eff}} \), there are two edge states on each edge. In addition to \(\bm{\Psi}_{l,1}\) and \(\bm{\Psi}_{r,1}\), the other two edge states are \(\bm{\Psi}_{l,2}\) and \(\bm{\Psi}_{r,2}\). The explicit forms of \(\bm{\Psi}_{l,2}\) and \(\bm{\Psi}_{r,2}\) are similar to Eq.~\eqref{edge_state}, with details provided in Appendix A of the End Matter.

\begin{figure}
\centering
\includegraphics[width=1\linewidth]{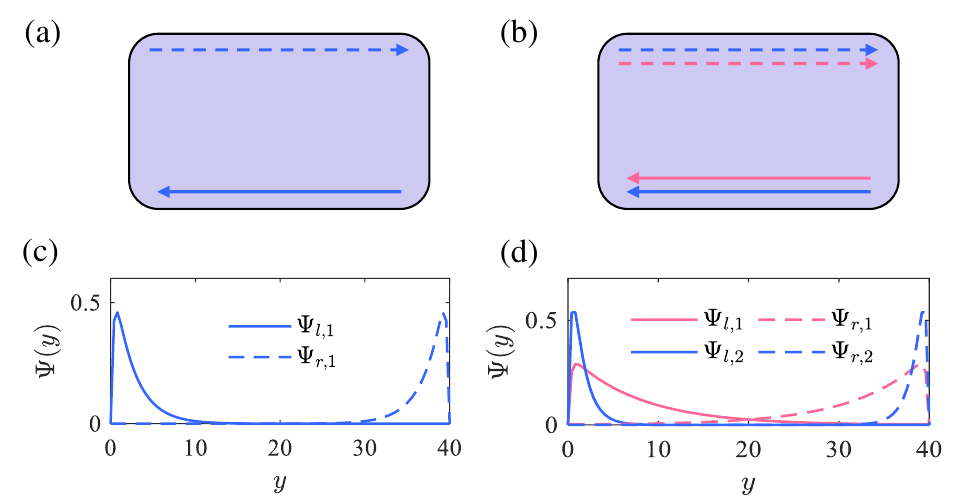}
\caption{(a,b) A schematic illustration of edge state distribution in the $x-y$ plane. The edge states near $y=0$ propagate to the $+x$ direction, and the edge states near $y=L_y$ propagate to the $-x$ direction. (c,d) Edge state distribution in the \(y\) direction, which is plotted by the analytical solution of wave function $\bm{\Psi}$. Only the first component of $\bm{\Psi}$ is plotted. The parameters are \(\Delta_{\text{eff}}=1\), \(A=B=1\) and \(L_y=N_ya=40\) with $a=1$. The system exhibits one edge states ($\mathcal{N}=1$) on each edge when \(m=-0.5\) (\(|m|<\Delta_{\text{eff}}\)), as illustrated in panels (a) and (c). In contrast, it displays two edge state ($\mathcal{N}=2$) on each edge for \(m=-1.5\) (\(m<-\Delta_{\text{eff}}\)), as shown in panels (b) and (d). The edge states localized in left (right) are labeled by solid (dashed) lines.  }
\label{fig:edges}
\end{figure}

When $k_{x}\neq0$ ($k_{x}$ is small), we employ degenerate perturbation theory by taking the perturbation Hamiltonian as: $\mathcal{H}_{1}(k_{x})= \mathcal{H}(k_{x},i\xi)-\mathcal{H}(0,i\xi)$.  As \( |m| < \Delta_{\text{eff}} \), in the degenerate space spanned by $ \{\bm{\Psi}_{l,1},\bm{\Psi}_{r,1}\}$, the perturbation Hamiltonian is diagonal: 
\begin{equation}
H_{1} \dot{=} Z \begin{bmatrix}
Ak_x & 0 \\
0 & -Ak_x
\end{bmatrix}.  
\end{equation}
Here, $Z=\Delta_s/(\Delta_s+\Delta_{\text{eff}})$ is the renormalization factor. The energy spectrum is linear in  $k_{x}$, indicating a massless edge quasi-particle with renormalized group velocity $ZA/\hbar$. The edge state near \( y = 0 \) (\( \bm{\Psi}_{l,1} \)) propagates exclusively in the \( +x \) direction, while the edge state near \( y=L_y \) (\( \bm{\Psi}_{r,1} \)) propagates exclusively in the \( -x \) direction [see Fig.~\ref{fig:edges}(a,c)]. This behavior is characteristic of chirality \cite{Read&Green2000}. The case for \( m < -\Delta_{\text{eff}} \) can be analyzed similarly, with the corresponding results shown in Fig.~\ref{fig:edges}(b,d).

Without crossing a phase boundary given by Eq.~(\ref{eq:Topo_condition}), the number of edge states remains the same. Therefore, we can extrapolate the results for $\mu_{\text{eff}}=0$ to obtain the distribution of edge states across the entire parameter space:

\begin{equation}
\mathcal{N} = 
\begin{cases} 
0, & m > \sqrt{\Delta_{\text{eff}}^{2} + \mu_{\text{eff}}^{2}}, \\ 
1, & |m| < \sqrt{\Delta_{\text{eff}}^{2} + \mu_{\text{eff}}^{2}}, \\ 
2, & m < -\sqrt{\Delta_{\text{eff}}^{2} + \mu_{\text{eff}}^{2}}.
\end{cases}
\label{eq:Topo_condition_mu_eff}
\end{equation}
This result is consistent with the the topological phase region determined by computing the Chern number, which captures the bulk-edge correspondence \cite{Hasan&Kane_2010}. It seems the same as the outcome obtained earlier in Refs.~\citep{Qi_2010,Chung_2011}, where the phase boundary is defined as $m^{2} = \Delta^{2} + \mu^{2}$. Notably, in our findings, $\mu_{\text{eff}}$ and $\Delta_{\text{eff}}$ are not independent parameters, as they both depend on the proximity coupling strength $ T $. Additionally, a significant shift $\mathrm{Re}(\chi)$ in the chemical potential is observed, which increases with proximity coupling strength. 

\begin{figure}
  \centering
\includegraphics[width=1\linewidth]{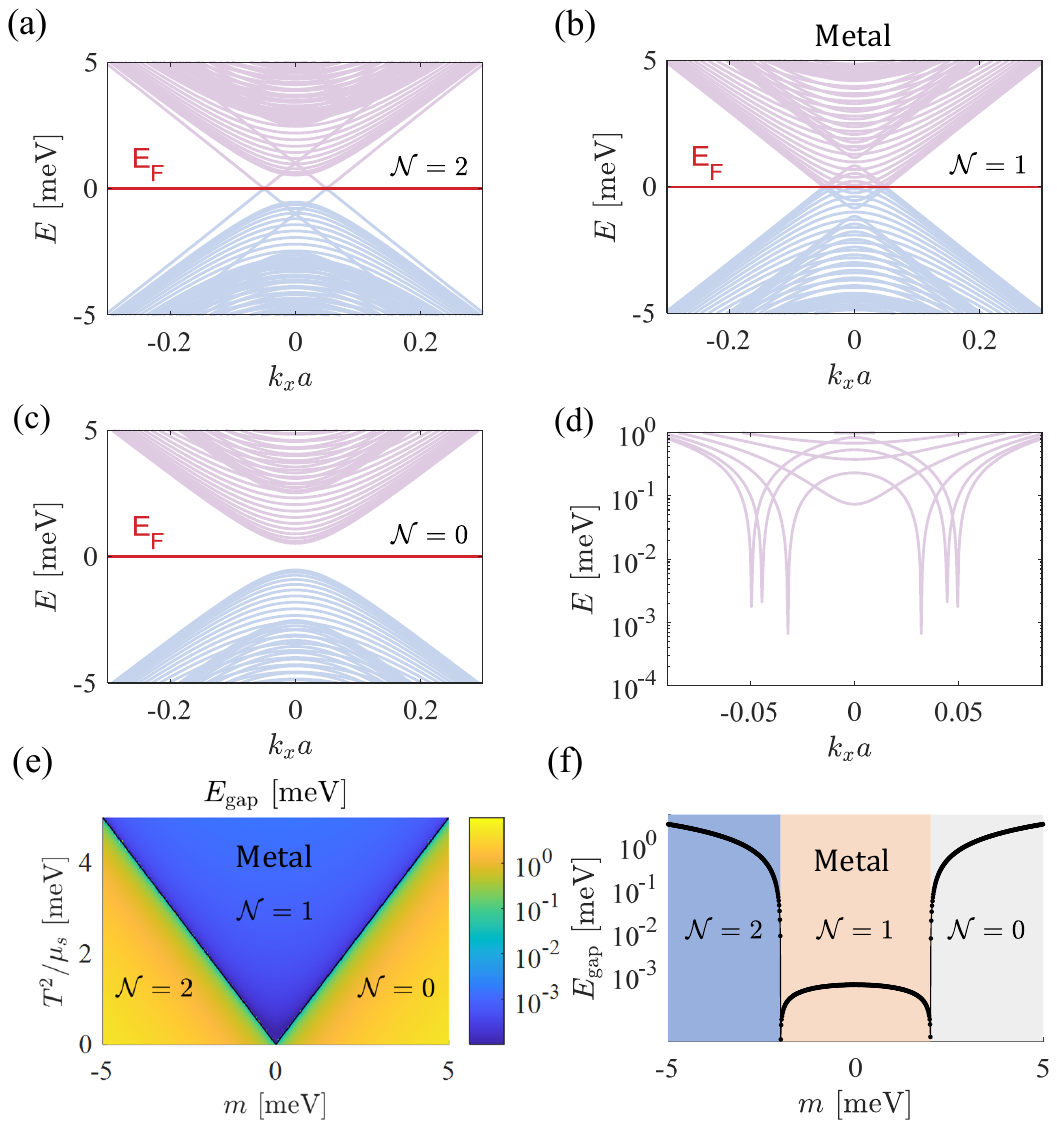}
 \caption{Energy spectrum of the holistic Hamiltonian with periodic boundary conditions in the $x$ direction and open boundary conditions in the $y$ direction. The parameters are set as follows: $\mu_m = 0$, $\mu_s = 5 \, \text{eV}$, $m_s = 9.1\times10^{-31} \, \text{kg}$, $\Delta_s = 1.5 \, \text{meV}$, $T=100 \,\text{meV}$, $A = 3 \, \text{eV\AA}$, and $B = 15 \, \text{eV\AA}^2$, with a lattice constant $a = 7.5 \, \mathrm{nm}$ and a length $L_y = N_y a = 1.5 \, \mathrm{\mu m}$ along the $y$ direction. Subplots (a-c) depict different values of the parameter $m$: (a) $m = -3 \, \text{meV}$, (b) $m = 0 \, \text{meV}$, and (c) $m = 3 \, \text{meV}$, corresponding to regions $\mathcal{N}=2$, $\mathcal{N}=1$, and $\mathcal{N}=0$, respectively. (b) illustrates the metallization effect, while (d) is an enlarged view of (b), highlighting the tiny gap at the micro-electronvolt scale. (e) Energy gap as a function of \(T\) and \(m\), denoted as \(E_{\text{gap}}(m, T)\). The color scale represents the $E_{\text{gap}}$ in meV, with regions labeled $\mathcal{N}=0, 1, 2$ indicating different topological phases. (f) Energy gap as a function of \(m\) under the condition \(T = 100 \, \text{meV}\). The color represent different topological regeion. The parameters used in (e) and (f) are the same as those in (a). }
    \label{fig2}
\end{figure}

\vspace{0.5em}
\emph{Metallization from chemical potential shift}--According to band thoery, metal is defined when the Fermi level situated in band and the system is gapless or the gap is very tiny. Therefore, when the chemical potential is shifted, the system may be render into metal. To illustate metallization effect caused by the shift of the chemical potential, we computate the engergy spectrum of holistical model (\ref{Holistical_BdG}). Specifically, we discretize the continuum model onto a square lattice model, and then compute the energy bands with periodic (open) boundary along the $x$ ($y$) directon \citep{Qi_2010}. Considering the practical parameters of the QAH-SC heterostructure, \(\mu_m = 0\, \text{meV}\), \(\mu_s = 5 \, \text{eV}\), \(\Delta_s = 1.5 \, \text{meV}\), \(T = 100 \, \text{meV}\), \(A = 3 \, \text{eV\,\AA}\), and \(B = 15 \, \text{eV\, \AA}^2\) \cite{2018PRB_KTlaw,NaturePhysics2010,SintificReport2023}, the computing results are presented in Fig.~\ref{fig2}(a-d). As illustrated in Fig.~\ref{fig2}(a), when \(\mathcal{N}=2\), two edge states are well-protected by the bulk gap. As \(m\) enters into the \(\mathcal{N}=1\) region, the bulk gap becomes very small, only a few microvolts, and the edge mode is submerged by the bulk state, as shown in Fig.~\ref{fig2}(b, d). Consequently, the hybrid system behaves like a metal. When \(\mathcal{N}=0\), the gap is large with no edge state, indicating a normal insulator state [see Fig.~\ref{fig2}(c)].

To visualize the metallization region, other parameters are set constant while \(T\) and \(m\) are varied. The band gap \(E_{\text{gap}}(T, m)\) is obtained numerically by identifying the minimum energy point of the model (\ref{Holistical_BdG}) with periodic boundary conditions applied in both the \(x\) and \(y\) directions. It can be seen from Fig.~\ref{fig2} (e,f) that the energy gap is nearly zero within the region \(\mathcal{N}=1\) (\(|m| < T^2 /\sqrt{\Delta_{s}^{2} + \mu_{s}^{2}}\)). Therefore, the metallization regions almost correspond to the \(\mathcal{N}=1\) region, which might also cause the half-integer signature, thus hindering the definitive identification of chiral Majorana fermions.

\vspace{0.5em}
\emph{Crossover from 2D superconductor to 3D}--In this section, we study the dependence of metallization on the thickness of SC so that the above 2D approach for chiral Majorana fermion transfer to a more realistic 3D system. We replace the momentum of Eq. (4) as $\mathbf{k}\rightarrow({\mathbf{k},k_z}) $. Here, $k_z=n\pi/d,n=1,2,3,\cdots$ is the wavenumber of standing wave in the $z$ direction. Besides, the proximity tunneling between the 2D QAH insulator and the 3D SC is:
\begin{equation}
H_{\mathrm{T}} =  \sum_{k_z}  T_{k_z} \int \frac{d^2k}{(2\pi)^2} \sum_{\sigma=\uparrow,\downarrow} 
\left[\varphi_{\mathbf{k}\sigma}^\dagger c_{\mathbf{k},k_z\sigma}  + \text{H.c.}\right].
\label{eq:HT3D}
\end{equation}
with $T_{k_z}=\sqrt{2/d}\int_0^d dz  {T(z)} \sin(k_zz).$

\begin{figure}
\includegraphics[width=1\linewidth]{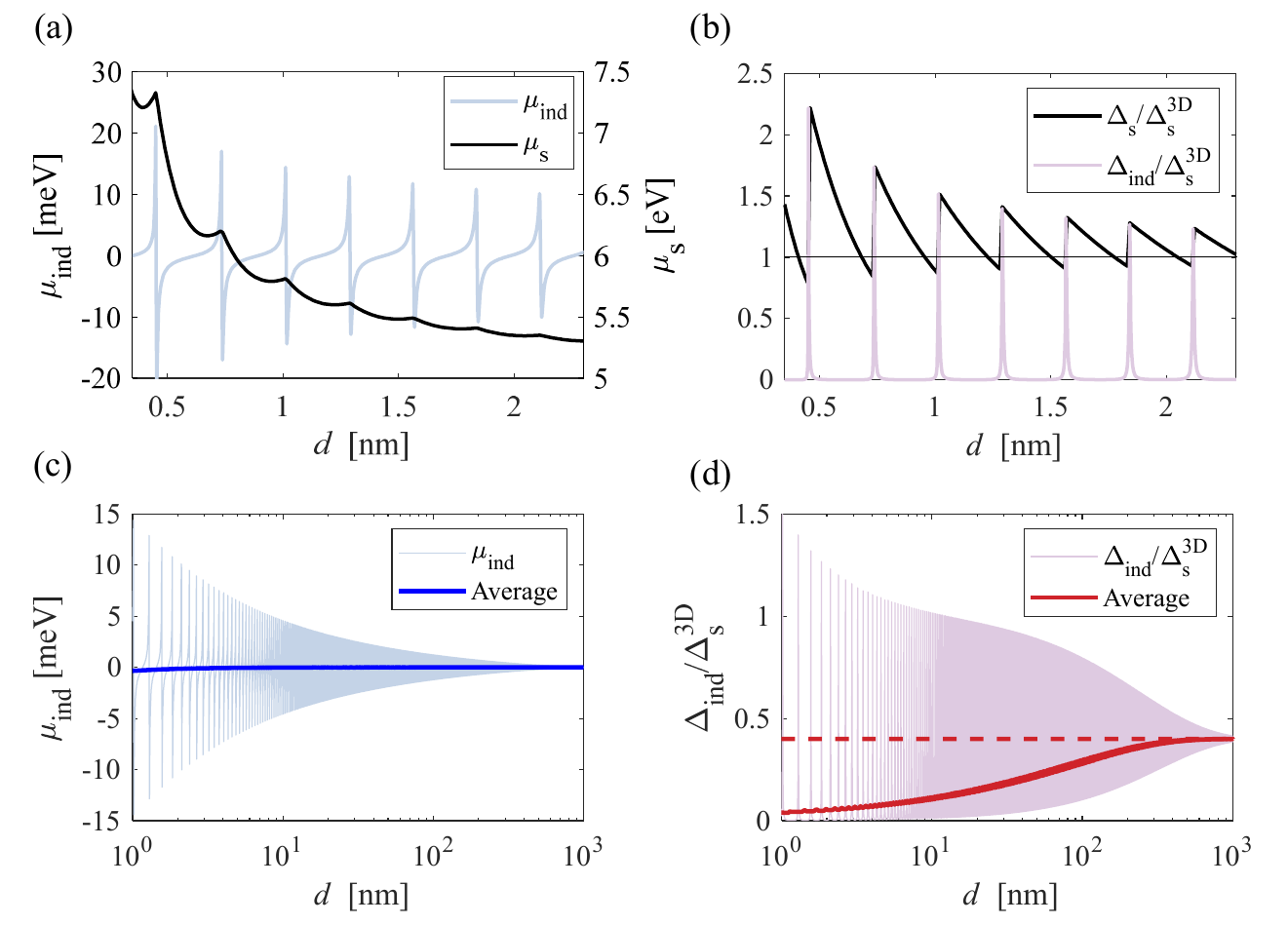}
    \caption{ (a) The supercondutor's chemical potential $\mu_s$ and induced chemical potential $\mu_{\text{ind}}$ versus thickness of superconductor layer $d$. (b) The supercondutor's engergy gap  $\Delta_s$ and induced engergy gap $\Delta_{\text{ind}}$ versus $d$. The intrinsic qunatities converge to the bulk value $\mu_s ^{\text{3D}}, \Delta_s ^{\text{3D}}$ soon while the induced qunatities continue to ossilate.  (c) The induced chemical potential $\mu_{\text{ind}}$ versus $d$  for wider range (light blue line). The dark blue line is the average of the light blue line. (b) Proximity-induced gap $\Delta_{\text{ind}}$ versus $d$ for wider range (light red line). The dark red line is an average of light red line. The dashed line is the analytical results in 3D limit when the superconductor is thick enough. The parameters are set as: $\mu_m = 0$, $\mu_s ^{\text{3D}} = 5 \, \text{eV}$, $m_s = 9.1\times10^{-31} \, \text{kg}$, $\hbar \omega_D=100K$, $\Delta_s^{\text{3D}} = 1.5 \, \text{meV}$, $T_0^2/k_F = 20000 \, [\text{meV}]^2$, $\kappa =k_F $ .  }
    \label{fig3_crossover}
\end{figure}

Similarly, the topological phase boundary is obtained as [see Appendix B for details] 
\begin{equation}
m^2 = \Delta_{\text{eff}}^2 + \mu_{\text{eff}}^2,
\label{Topo_condition3D}
\end{equation}
where $\mu_{\text{eff}} = \mu_m + \mathrm{Re}(\chi)$ and $\Delta_{\text{eff}} = \mathrm{Im}(\chi)$. The effective shfit of the chemical potential and pairing strength depends on contributions from all subbands indicated by $k_z$:
\begin{equation}
\chi = \sum_{k_z}  \frac{|T_{k_z}|^2}{\epsilon(k_z) - i \Delta_s}, \quad \epsilon(k_z) = \frac{\hbar^2 k_z^2}{2 m_s}  - \mu_s. 
\label{shift}
\end{equation}

It follows from Eq.~(\ref{Topo_condition3D}) that $\mu_{\text{eff}}$ and $\Delta_{\text{eff}}$ together determine the topological transition condition. Furthermore, they can directly establish a connection with observable quantities (the proximity-induced chemical potential shift $\mu_{\text{ind}}$ and proximity-induced gap $\Delta_{\text{ind}}$) through a low-energy effective Hamiltonian [see Appendix C]: $\mu_{\text{ind}} = Z\mu_{\text{eff}}$ and $\Delta_{\text{ind}} = Z\Delta_{\text{eff}}$  \cite{2010Proximity,2017lowEnergy_renormalization}. Here, $Z = \Delta_s/(\Delta_s + \Delta_{\text{eff}})$ is the renormalization factor, and $\Delta_{\text{eff}}$ characterizes the effective coupling strength: in the strong coupling regime ($\Delta_{\text{eff}} \gg \Delta_s$), $\Delta_{\text{ind}} \to \Delta_s$, while for weak coupling ($\Delta_{\text{eff}} \ll \Delta_s$), $\Delta_{\text{ind}} \approx \Delta_{\text{eff}}$.

For thin-film superconductors, their energy gap and chemical potential oscillate as a function of film thickness \cite{1963_Blatt_Thompson, Thompson1963,
1984PhysRevLett.53.2046, 2004Science_Xue}, i.e., $\Delta_s = \Delta_s(d)$ and $\mu_s = \mu_s(d)$. These oscillations, as illustrated in Figs.~\ref{fig3_crossover}(a,b), are obtained by self-consistently \cite{1966DeGennes,2016BdG} solving the multi-band energy-gap equation under open boundary conditions \cite{1963_Blatt_Thompson,Thompson1963,2020shape_resonances} [See Appendix D]. For an exponentially decaying coupling strength $T(z) = T_0 \exp(-\kappa z)$, where $\kappa^{-1}$ is the characteristic decay length, we compute the induced chemical potential $\mu_{\text{ind}}$ and energy gap $\Delta_{\text{ind}}$ as functions of the SC thickness $d$, shown in Fig.~\ref{fig3_crossover}.

It is worth noting that the oscillations of the induced quantities $\Delta_{\text{ind}}, \mu_{\text{ind}}$ and the intrinsic superconducting quantities $\Delta_s, \mu_s$ share the same periodicity, determined by the superconductor's Fermi wavelength $\lambda_F$. Resonance peaks occur when one subband crosses the Fermi surface at $d = n\lambda_F/2$, where $n$ is an integer. However, these two exhibit distinct characteristic length scales. The oscillations of the intrinsic superconducting quantities virtually vanish when $d \gg \eta = \frac{3}{8} \lambda_F/\lambda$, where $\lambda = g(0)V$ is the dimensionless superconducting coupling constant [see Appendix D]. Due to the inverse proportional decay of the oscillations \cite{Thompson1963}, these become negligible after a few dozen Fermi wavelengths. In contrast, the characteristic length scale for the oscillations of the induced quantities is the coherence length \cite{2017_Finite}, and due to the exponential decay of oscillations \cite{qiao2025sizeoptimizationobserveingmajorana}, the resonance effects vanish after a few coherence lengths. Consequently, the summation over $k_z$ in Eq.~\eqref{shift} transitions to an integral, resulting in $\mu_{\text{eff}} \simeq 0$ and $\Delta_{\text{eff}} = (T^2/\hbar)\sqrt{m_s/2\mu_s}$ \cite{Alicea_2012,Yue2023}, with $T = T_{k_F}\sqrt{d}$, as depicted by the dashed line in Fig.~\ref{fig3_crossover}(d).

For the superconducting material niobium, commonly used in experiments aimed at detecting chiral Majorana fermions \cite{He2017chiral,2020Absence,Huang_2024,Uday2024}, within the electron gas approximation, the characeristic length are calculated as $\eta \approx 0.5$~nm and $\xi_s=\hbar v_F/\pi\Delta_s^{\text{3D}} \approx 180$~nm\footnote{It should be note that this value of coherence length deviates from the experimental measured value of $\xi_s \approx 40$~nm \cite{2023Nb,2025niobium} since the real elctron structure of niobium deviates from the free electron gas approximation with high degree of anisotropy \cite{1987Nb_anisotropy}}. When $d > 20\eta$, $\Delta_s$ and $\mu_s$ can be approximated as constants with less than 5\% error; When $d > 5\xi_s$, $\Delta_{\text{ind}}$ becomes approximately constant with less than 5\% error. Notably, thin superconducting layers at resonance can induce gaps larger than those observed in the thick-SC limit. Such enhanced energy gaps can extend the range for observing signatures of chiral Majorana fermions (\(0.5 \, e^2/h\)), as discussed below.

Remarkably, thin superconducting layers at resonance can induce gaps larger than those observed in the thick-SC limit. Such enhanced energy gaps can extend the range for observing signatures of chiral Majorana fermions (\( 0.5 \, e^2/h \)), as discussed below.

\begin{figure}
 \centering  \includegraphics[width=1\linewidth]{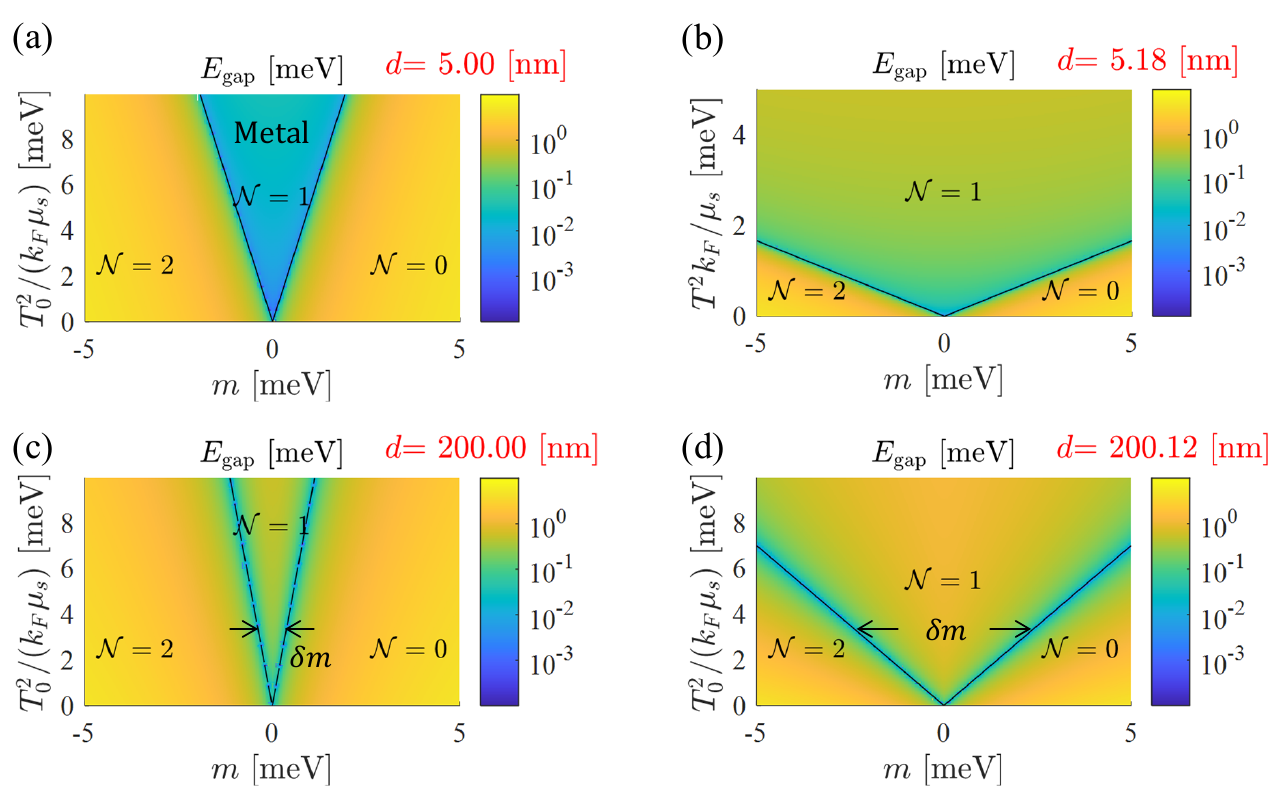}
    \caption{
  Energy gap diagrams for various thicknesses $d$ (a) $d = 5.00$\,nm, (b) $d = 5.18$\,nm, (c) $d = 200.00$\,nm, and (d) $d = 200.12$\,nm. Other parameters are set to be: $\mu_m = 0$, $\mu_s = 5 \, \text{eV}$, $m_s = 9.1\times10^{-31} \, \text{kg}$, $\Delta_s = 1.5 \, \text{meV}$, $A = 3 \, \text{eV\AA}$,  $B = 15 \, \text{eV\AA}^2$,  and $\kappa =k_F $.}
    \label{fig:Fig4_metallization_disappear}
\end{figure}

To visualize the width of this window for observing chiral Majorana fermion signatures, we compute the energy gap and the phase diagram as a function of superconductor thickness \( d \) (Fig.~\ref{fig:Fig4_metallization_disappear}). At \( d = 5 \, \mathrm{nm} \), the phase diagram closely resembles the 2D SC case [Fig.~\ref{fig:Fig4_metallization_disappear}(a) and 3(e)], where the \(\mathcal{N}=1\) region exhibits a tiny energy gap (\(\sim \mu\mathrm{eV}\)) and behaves metallically. When \( d = 5.18 \, \mathrm{nm} \) (near the resonance point \( d \simeq 19\lambda_F/2 \)), the induced gap approaches the SC bulk gap, shrinking the metallization region near phase boundaries [Fig.~\ref{fig:Fig4_metallization_disappear}(b)].

The non-resonant thickness $ d = 200.00 \, \mathrm{nm} $ produces a narrower signature window compared to the resonant case $ d = 5.18 \, \mathrm{nm} $ [Fig.~\ref{fig:Fig4_metallization_disappear}(b,c)]. Remarkably, a minor adjustment to $ d = 200.12 \, \mathrm{nm} $ (near $ 732\lambda_F/2 $) creates a broad signature window approximately 7 times wider than at $ d = 200.00 \, \mathrm{nm} $ [Fig.~\ref{fig:Fig4_metallization_disappear}(c,d)]. This sensitivity to thickness variations can be explanied from Eq.~\eqref{Topo_condition3D}: The boundary condition $ m = \pm \sqrt{\Delta_{\text{eff}}^{2} + \mu_{\text{eff}}^{2}} $ determines the $\mathcal{N}=1$ window width as $ \delta m = 2\sqrt{\Delta_{\rm eff}^2 + \mu_{\rm eff}^2} $. Because both $\Delta_{\rm eff}$ and $\mu_{\rm eff}$ oscillate with $ d $, the resulting $\delta m$ exhibits strong thickness dependence.

\vspace{0.5em}  
\emph{Conclusion and Discussion}--We developed a microscopic theory for chiral Majorana fermions, which is suitable for arbitrary proximity coupling strength. We also systematically investigate how the finite thickness of the superconductor (SC) affects the detection of the chiral Majorana fermion. There are three important findings: (i) Periodic structure of metallization. For thin SCs ($\sim$10\,nm), the metallization region oscillates with the thickness of the superconductor, and its period corresponds to the Fermi wavelength of SC ($\lambda_F$). Metallization occurs except near resonance thicknesses where $d \approx n\lambda_F/2$. (ii) Periodic structure of chiral Majorana fermions. For intermediate thicknesses ($\sim$100\,nm), the window widtth for observing chiral Majorana fermions changes periodically with the thickness of the superconductor. The oscillation period of this window is also $\lambda_F$. (iii) Stable structure of chiral Majorana fermions. For thick SCs ($\sim$1000\,nm),  the chiral Majorana fermion shows uniform behavior as $d$ varies. Metallization is absent in both (ii) and (iii). All these predictions regarding the thickness dependence can be experimentally verified.

Crucially, near resonance points (\( d \approx n\lambda_F/2 \)), the induced gap can reach the SC gap, expanding the \(\mathcal{N}=1\) topological phase. This widens the mass gap range for observing the signature of chiral Majorana fermion (\( 0.5 \, e^2/h \)). Given that experiments report half-integer plateaus only in narrow mass gap intervals, precise control of \( d \) could enhance the detectable range. Optimizing \( d \) may thus improve data quality and provide clearer evidence for chiral Majorana fermions.  

Our analysis identifies two metallization criteria: substantial chemical potential shifts and ultra-small induced gaps ($\Delta_{\text{ind}}\sim\mu$eV). While the model handles arbitrary coupling strengths, unexplained metallization in strong-coupling regimes implies missing physics--likely from chemical potential inhomogeneity, interface structure, or non-ideal tunneling. These findings provide concrete guidelines for observing chiral Majorana fermion while highlighting the need for refined models of realistic heterostructures.

\vspace{0.5em}
\emph{Acknowledgments}--The authors appreciate quite much for the helpful discussion with Zhi-Lei Zhang and R. X. Zhai. This study is supported by the National Natural Science Foundation of China (NSFC) (Grant No. 12088101) and NSAF (Grants No. U2330401).
\bibliographystyle{apsrev4-2}
\bibliography{Refs}

\onecolumngrid

\begin{center}
  \textbf{\LARGE End Matter}
\end{center}

\twocolumngrid

\vspace{0.5em}
\emph{Appendix A}--For the specific case where $\mu_{\text{eff}}=0$ and under the approximation that $ \mu_s +\hbar^2\xi^2/(2m_s)\approx\mu_s$, the determinant can be factored into the product of four terms: 
$
\text{det}[\mathcal{H}(0,i\xi)] =F_{1}F_{2}F_{3}F_{4},
$  
where
$F_1 = B\xi^2 - A\xi + \Delta_{\text{eff}} - m,\, F_2 = B\xi^2 - A\xi - \Delta_{\text{eff}} - m,\, F_3 = B\xi^2 + A\xi + \Delta_{\text{eff}} - m,\, F_4 = B\xi^2 + A\xi - \Delta_{\text{eff}} - m.$ 

Each quadratic corresponds to an equation in terms of \(\xi\), whose solutions are:
\[
\begin{aligned}
\xi_{\pm,1} = \frac{A \pm \sqrt{A^2 - 4B(\Delta_{\text{eff}} - m)}}{2B}, \quad
\xi_{\pm,3} = -\xi_{\mp,1} \\
\xi_{\pm,2} = \frac{A \pm \sqrt{A^2 + 4B(\Delta_{\text{eff}} + m)}}{2B}, \quad 
\xi_{\pm,4} = -\xi_{\mp,2}
\end{aligned}
\]
Each \(\xi\) is substituted into the eigenvalue equation to solve for specific values of the superposition coefficients \(\bm{u}, \bm{v}\). By satisfying the open boundary conditions \(\bm{\Psi}(0) = \bm{\Psi}(L_y) = 0\), each quadratic provides one solution, yielding a total of four solutions (two of them are alredy given in the text).
\[
\begin{aligned}
\bm{\Psi}_{r,2}(y) & \propto \begin{pmatrix} \bm{u}_+ \\ -\bm{u}_+ \end{pmatrix} 
\big( e^{\xi_{+,2}(y-L_y)} - e^{\xi_{-,2}(y-L_y)} \big), \\
\bm{\Psi}_{l,2}(y) &\propto \begin{pmatrix} \bm{u}_- \\ \bm{u}_- \end{pmatrix} 
\big( e^{-\xi_{+,2}y} - e^{-\xi_{-,2}y} \big).
\end{aligned}
\]

The parameter domain in which $\bm{\Psi}_{r,1}(y)$ and $\bm{\Psi}_{l,1}(y)$ are valid is \( m < \Delta_{\text{eff}} \). This arises from the fact that when \( m > \Delta_{\text{eff}} \), the decay factors \( \xi_{+,1} \) and \( \xi_{-,1} \) have opposite signs, making it impossible to satisfy the boundary conditions simultaneously. Conversely, for \( m < \Delta_{\text{eff}} \), both \( \xi_{+,1} \) and \( \xi_{-,1} \) share the same sign (both being positive), allowing their linear combination to satisfy the boundary conditions at both ends consistently.

Similarly, for \(\bm{\Psi}_{r,2}(y)\) and \(\bm{\Psi}_{l,2}(y)\), the parameter domain in which they are valid is \( m < -\Delta_{\text{eff}} \). When \( m > \Delta_{\text{eff}} \), \( \xi_{+,1} \) and \( \xi_{-,1} \) have opposite signs, as do \( \xi_{+,2} \) and \( \xi_{-,2} \), meaning that no valid solution can be constructed under these circumstances.

The results are summarized in the table below:

\[
\begin{array}{|c|c|c|c|c|c|c|}
\hline
 & \xi_{+,1} & \xi_{-,1} & \text{Solution} & \xi_{+,2} & \xi_{-,2} & \text{Solution} \\ \hline
m > \Delta_\text{eff} & >0 & <0 & \text{None} & >0 & <0 & \text{None} \\ \hline
|m| < \Delta_\text{eff} & >0 & >0 &   \bm{\Psi}_{r,1},  \bm{\Psi}_{l,1} & >0 & <0 & \text{None} \\ \hline
m < -\Delta_\text{eff} & >0 & >0 & \bm{\Psi}_{r,1}, \bm{\Psi}_{l,1} & >0 & >0 & \bm{\Psi}_{r,2}, \bm{\Psi}_{l,2} \\ \hline
\end{array}
\]

\vspace{0.5em}
\emph{Appendix B}--The tunning Hamilontian in real-space is:

\[ 
H_\text{T} =  \iint dxdydz 
 \, T(z)\sum_{\sigma=\uparrow, \downarrow}\left[\varphi_\sigma^{\dagger}(x, y) c_\sigma(x, y, z) + \text{H.c.}\right] 
\]
We apply a Fourier transformation to this Hamiltonian, representing the field operators in momentum space as follows:

\[
\varphi_{\sigma}(\mathbf{x}) = \int \frac{d^2 k}{(2 \pi)^2} \varphi_{\mathbf{k} \sigma} e^{i \mathbf{k} \cdot \mathbf{x}}, \,
c_{\sigma}(\mathbf{x},z) = \int \frac{d^2 k}{(2 \pi)^2} c_{\mathbf{k} \sigma}(z) e^{i \mathbf{k} \cdot \mathbf{x}}
\]
By substituting these expressions into the real-space Hamiltonian, we obtain the momentum-space representation:

\[ 
H_{\mathrm{T}} =  \iint \frac{d^2k}{(2\pi)^2} \int dz\, T(z)\sum_{\sigma=\uparrow, \downarrow} \left[ \varphi_{\mathbf{k}\sigma}^\dagger c_{\mathbf{k}\sigma}(z) + \text{H.c.}\right] 
\]

To satisify the open boundary condition, the eigenfunctions in the 
 $z$-direction take the form of standing waves $\sin(k_z z)$, resulting a real Fourier transformation:
\[
c_{\mathbf{k}\sigma}(z) = \sqrt{\frac{2}{d}} \sum_{k_z} \sin\left({k_z z}\right) c_{\mathbf{k}\sigma,k_z}, \quad \text{with  } \quad k_z = \frac{\pi n}{d}.
\]
 
 Consequently, the Hamiltonian becomes:

\[ 
H_{\mathrm{T}} = \sum_{k_z} T_{k_z} \int \frac{d^2k}{(2\pi)^2} \sum_{\sigma=\uparrow, \downarrow} \left[\varphi_{\mathbf{k}\sigma}^\dagger c_{\mathbf{k}\sigma,k_z} + \text{H.c.}\right] 
\]
with 
\[T_{k_z}=\sqrt{\frac{2}{d}}\int_0^d dz  {T(z)} \sin(k_zz).\]

For an exponentially decaying coupling strength $T(z) = T_0 e^{-\kappa z}$ (where $\kappa^{-1}$ is the characteristic decay length), the integral is:
\[
T_{k_z}= \sqrt{\frac{2}{d}} \frac{T_0 k_z}{\kappa^2 + k_z^2} \left[ 1 - (-1)^n e^{-\kappa d} \right].
\]

We consider the regime where $\kappa d \gg 1$, which results in
\(
T_{k_z} \approx \sqrt{\frac{2}{d}} \frac{T_0 k_z}{\kappa^2 + k_z^2}.
\label{eq:strong_coupling}
\)

In the basis $[\boldsymbol{\varphi}_{\mathbf{k}},\boldsymbol{\varphi}_{-\mathbf{k}}^{\dagger},
\boldsymbol{C}_{\mathbf{k}, k_{z_1}}     ,
\boldsymbol{C}_{\mathbf{k}, k_{z_2}}
,...,\boldsymbol{C}_{\mathbf{k}, k_{z_N}}   ]^{T}$, where each component $\boldsymbol{C}_{\mathbf{k}, k_z}$ is defined as $[\boldsymbol{c}_{\mathbf{k},k_{z}},\boldsymbol{c}_{-\mathbf{k},k_{z}}^{\dagger} ]^{T} $, the holistic Hamiltonian for the heterostructure is expressed as a block matrix:

\[
\mathcal{H}(\mathbf{k}) = \begin{pmatrix} \mathcal{H}_{\text{Q}}(\mathbf{k}) & \mathcal{T} \\ \mathcal{T}^{\dagger} & \mathcal{H}_{\text{SC}} (\mathbf{k})\end{pmatrix}.
\]
In this representation, \(\mathcal{H}_{\text{Q}}\) is the Hamiltonian block coppresponding to the quantum anomalous Hall insulator, \(\mathcal{H}_{\text{SC}}\)  represents the Hamiltonian block for the superconductor, and \(\mathcal{T}\) signifies the coupling between them. Similiar to the 2D case in the main text, the condition \(
\text{det}[\mathcal{H}(\mathbf{k}=0)] = 0
\) determines the phase boundary at which the holistic edge mode emerges or disappears. Using Schur Complement formula, the determinant can be decomposed as
\(
\text{det}(\mathcal{H}) = \text{det}(\mathcal{H}_{\text{SC}}) \cdot \text{det}(\mathcal{H}_{\text{Schur}})
\), where the Schur complement, \(\mathcal{H}_{\text{Schur}} \equiv \mathcal{H}_{\text{Q}} - \mathcal{T} \mathcal{H}_{\text{SC}}^{-1} \mathcal{T}^{\dagger}\). Given that the determinant of the superconductor block satisfies \(\text{det}(\mathcal{H}_{\text{SC}}) > 0\), the topological condition \(\text{det}(\mathcal{H}) = 0\) equates to 
\(
\text{det}(\mathcal{H}_{\text{Schur}}) = 0.
\) Through matrix algebra calculations, the Schur complement \(\mathcal{H}_{\text{Schur}}(\mathbf{k})\) is:

\[
\mathcal{H}_{\text{Schur}}(\mathbf{k}=0)  = \begin{pmatrix}
m\sigma_z-\mu_{\text{eff}}  & i \Delta_{\text{eff}}\sigma_y \\
-i \Delta_{\text{eff}} \sigma_y & -m\sigma_z+\mu_{\text{eff}}
\end{pmatrix},
\]
where the effective chemical potential and the effective superconducting gap are \(\mu_{\text{eff}} = \mu_m + \mathrm{Re}(\chi)\) and \(\Delta_{\text{eff}} = \mathrm{Im}(\chi)\). These expressions rely on the auxiliary quantity \(\chi\), given by:

\[
\chi = \sum_{k_z} \frac{|T_{k_z}|^2}{\epsilon(k_z) - i \Delta_s}, \quad \epsilon(k_z) = \frac{\hbar^2}{2 m_s} k_z^2 - \mu_s.
\]
Then the topological transition condition is obtained by \(\text{det}[\mathcal{H}_{\text{Schur}}(\mathbf{k}=0)] = 0\), which is Eq.~(13) in the main text.

\vspace{0.5em}
\emph{Appendix C}--Usually, the low-energy effective Hamiltonian and the renormalization factor are derived using the Green's function method \cite{2010Proximity,2017lowEnergy_renormalization}. Here, we propose an alternative approach based on the elimination method.
What is more, we extend this approach to be applicable to finite size thickness $d$.

The eigenvalue equation for the system is given by:

\[
 \begin{pmatrix}
\mathcal{H}_{\text{Q}} & \mathcal{T} \\
\mathcal{T}^{\dagger} & \mathcal{H}_{\text{SC}}
\end{pmatrix}
\begin{pmatrix}
\psi_m  \\
\psi_s 
\end{pmatrix} = E \begin{pmatrix}
\psi_m  \\
\psi_s 
\end{pmatrix}.
\]
Eliminating \(\psi_s\), we obtain the following equation in terms of \(\psi_m\) and \(E\):

\[
(\mathcal{H}_{\text{Q}} + \mathcal{T}(E - \mathcal{H}_{\text{SC}})^{-1} \mathcal{T}^{\dagger})\psi_m = E\psi_m.
\]
In the low-energy limit, where \(E/\Delta_s \ll 1\), the energy \(E\) can be treated as a perturbation. Using a Taylor expansion, one obtains the following:

\[
\left(\mathbbm{1}  + \mathcal{T}\mathcal{H}_{\text{SC}}^{-2}\mathcal{T}^{\dagger}\right)^{-1}(\mathcal{H}_{\text{Q}} - \mathcal{T} \mathcal{H}_{\text{SC}}^{-1} \mathcal{T}^{\dagger})\psi_m = E\psi_m.
\]
Thus, the low-energy effective Hamiltonian is expressed as:
\[
\begin{aligned}
\mathcal{H}_{\text{L}} (\mathbf{k}) 
&= \left( \mathbbm{1} + \mathcal{T}\mathcal{H}_{\text{SC}}^{-2} \mathcal{T}^{\dagger} \right)^{-1} \left( \mathcal{H}_{\text{Q}} - \mathcal{T} \mathcal{H}_{\text{SC}}^{-1} \mathcal{T}^{\dagger} \right)\\
&= Z
\begin{pmatrix}
h_{\text{Q}}(\mathbf{k}) - \mu_{\text{eff}}  & i \Delta_{\text{eff}} \sigma_y \\
-i \Delta_{\text{eff}} \sigma_y & -h_{\text{Q}}^*(-\mathbf{k}) + \mu_{\text{eff}}
\end{pmatrix}.
\end{aligned}
\]

Here, the renormalization factor \(Z\) is defined as \( Z = \Delta_s / (\Delta_s + \Delta_{\text{eff}}) \). Hence, it can be observed that the proximity-induced chemical potentail shift $\mu_{\text{ind}}$ and energy gap $\Delta_{\text{ind}}$ are given by $\mu_{\text{ind}} = Z\mu_{\text{eff}}$ and $\Delta_{\text{ind}} = Z\Delta_{\text{eff}}$, respectively.

\vspace{0.5em}
\emph{Appendix D}--We analyze thickness-dependent superconductivity in a thin film where electron quantization yields subbands 
\(
\epsilon_n = \frac{\hbar^2}{2m}\left(\frac{n\pi}{d}\right)^2.\) When the $n$-th subband touches the Fermi level ($\mu=\epsilon_n$), requiring fixed electron density $\rho_e$ gives the critical thickness:
\[
d_c(n) = \left[\frac{\pi}{2\rho_e}\left(n^3 - \frac{1}{6}n(n+1)(2n+1)\right)\right]^{1/3}.
\]

For \(d \in [d_c(n), d_c(n+1)]\), the Fermi level follows:
\[
\mu(d) = \frac{2\rho_e d^3/\pi + \frac{1}{6}n(n+1)(2n+1)}{n} \,
\frac{\hbar^2\pi^2}{2md^2}.
\]

Using the self-consistent method in real space (also known as the BdG method)~\cite{1966DeGennes,2016BdG}, the multi-band gap equation is derived \cite{1963_Blatt_Thompson,Thompson1963}:
\[
\Delta_k = \frac{1}{2}\sum_q V_{kq} \Delta_q \int g_{2D} \frac{\mathrm{d}\varepsilon}{\sqrt{\left(\varepsilon + \epsilon_q - \mu\right)^2 + \Delta_q^2}},
\]
where the inter-subband coupling is given by
\[
V_{kq} \equiv V \left(\sqrt{\frac{2}{d}} \right)^4 \int_0^d \mathrm{d}z \, \sin^2(kz)\sin^2(qz) = \frac{V}{d}\left(1 + \frac{\delta_{kq}}{2}\right).
\]

Assuming a uniform gap approximation $\Delta_k \approx \Delta$, we obtain:
\[
1 = g_{2D}V\,\frac{n + \frac{1}{2}}{2d} \int_{-\hbar\omega_D}^{\hbar\omega_D}
\frac{\mathrm{d}\varepsilon}{\sqrt{\varepsilon^2 + \Delta^2}},
\quad g_{2D} = \frac{m}{\pi\hbar^2}.
\]

The above equation can be solved explicitly:
\[
\Delta(d) = 2\hbar\omega_D\,
\exp\!\left[-\frac{1}{g_{2D}V}\,\frac{d}{n + \frac{1}{2}}\right],
\quad
d \in [d_c(n), d_c(n+1)].
\]

At subband occupation transitions, the gap exhibits resonance peaks, expressed as \cite{Thompson1963}:
\[
\Delta_{\mathrm{peak}}(n) = 2\hbar\omega_D\,
\exp\!\left[-\frac{1}{g_{2D}V}\,\frac{d_c(n)}{n + \frac{1}{2}}\right].
\]

Expanding \(\Delta_{\mathrm{peak}}(n)/\Delta_{\mathrm{bulk}}\) for large \(d\), we derive the thickness-dependent enhancement:
\[
\frac{\Delta_{\mathrm{peak}}(n)}{\Delta_{\mathrm{bulk}}} = 1 + \frac{\eta}{d},
\quad \eta = \frac{3\lambda_F}{8g_{3D}V}.
\]
Here, \(g_{3D} = \frac{m}{\pi^2 \hbar^2} k_F\) is the three-dimensional density of states at the Fermi surface.

\end{document}